**Brief Report**

# Co-evolution of Mercy and Altruistic Cooperation

**Wang zhijian and Luo Weidong**

**Interdiscriptional Center for Social Sciences, Zhejiang University, 310027, China**

## Abstract

**Besides altruistic punishment and group selection, we argue that, mercy can lead to altruistic cooperation. Modeling the micro economic behavior of the mercy, with two alleles of genes (Cooperation or Defection & Mercy or No mercy) agents in a network, we present the computational simulation results in the spatiotemporal evolution game theory frame to prove the above argument. Here, mercy (or as 'Love thy neighbors') means, the agents, with mercy preference, might share his own fitness with his poorest neighbor who poorer than himself.**

# 同情导致合作

王志坚 罗卫东[1]

ICSS, 浙江大学, 杭州, 310027 中国

**摘要：**利他惩罚和群体选择可以形成利他合作。本文说明，同情(或:爱你的邻居)一样可以导致合作。通过对同情行为的微观描述，在双等位基因个体组成群体的时空演化博弈框架下，我们通过计算机仿真结果证明上述结论。这份简报通过模型与仿真结果说明“同情导致合作”，同时说明同情的微观作用机制以及系统特性。本文中“同情(或:爱你的邻居)”指，当自己最穷的邻居适存度小于自己时候，有怜悯偏好者与自己最穷的邻居之间按一定比例分享适存度。

## 介绍:

除了利他惩罚[Boyd2007]和群体选择[Bowles2007]二条路径，本文证明，同情可以导致合作。本文中“同情(或:爱你的邻居)”指，当自己最穷的邻居适存度小于自己时候，有怜悯偏好者与自己最穷的邻居之间按一定比例分享适存度。我们通过对同情行为的微观描述, 在演化博弈论基础上，通过计算机仿真进行研究。 结果表明，与利他惩罚一样，同情可以强化利他合作的形成。在我们的模型中，与参考文献[Bowles2007]一样，我们假设每个人有二种等位的独立偏好。博弈过程参考了常用的带有利他惩罚的公共品演化博弈框架, 只是用“同情行为”替代了“利他惩罚”行为。同情的存在，同时使得网络结构中显示为动态的、丰富的、多样性的偏好共存。通过定量计算，本文说明具备同情行为的系统，会涌现出稳定的合作结构。我们的研究结果显示，合作结构的形成，无论在时间和空间上，都依赖于同情行为。一定程度上，本文的结果是对 “爱邻居如同爱自己”这个古老箴言中潜藏着的人类智慧的解读。

## 模型结构:

N×N 的二维正方形网格中[Nowak2006[2]]，每个格点上各有一人。 由描述合作的和描述同情行为的二种特性(基因)组成每个人偏好[Bowles2007[3]]。 每种特性(基因)偏好又分正负面，这样系统角色有四种: 既合作又有同情心的角色(CM■)，合作但是没有同情心的角色(CN

■), 不合作但是有同情心的角色(DM■), 以及不合作而且没有同情心的角色(DN■)。 以下文中，角色 C 表示含 C 基因的角色，D，M 和 N 也一样。

## 游戏规则:

游戏的步骤由以下几步组成: 1, 邻居公共品博弈；2, 同情行为（爱你的邻居)；3, 自然选择；4, 变异并生成新一代。

1, 邻居公共品博弈。每轮博弈中, 每人轮流做庄一次, 也就是说, 每轮博弈由 N×N 次子博弈组成。 每次子博弈是一个格点与其四近邻展开一次五人公共品博弈。 所有 C 偏好者, 包括(CM 和 CN)在每次子博弈中支付 $c = 1$，公共品的效益倍数为 b。

2, 爱你的邻居(同情行为)。所有格点再做庄一次展开子游戏。这次决定是否爱你的邻居。所有 M 偏好者 (包括 CM 和 DM) 在他自己做庄时候, 各自找到近邻中最穷者, 当自己最穷的邻居适存度小于自己时候，计算财富差值, 并按差值的一定比例(怜悯度)送他适存度(fitness)。

3, 自然选择。自然选择是淘汰适存度最低的人，淘汰比例为φ。被淘汰者的位子上的新一代人模仿四近邻中最高的适存度的人的偏好。

4, 变异。按照变异率新一代将自己的基因变换成其他的基因种类。这样新一代就形成了。

初始化系统后, 按照上述步骤让系统进行结算, 并重复。经过多次重复, 系统可以获得演化稳定结构的结果。

## 仿真结果:

(一) 有无同情行为的结果对比

系统完全随机初始化，其他条件完全相同。可以通过是否执行上述步骤2 的结果对比, 说明同情行为对合作形成的作用。执行步骤 2, 是“有同情行为”(见图一(b)), 省略执行步骤 2 是“无同情行为”(见图一(a))。实验表明, 该模型中, 同情行为对合作的生成起到决定性的作用。同情行为的存在，使得系统中 C 得以生存。

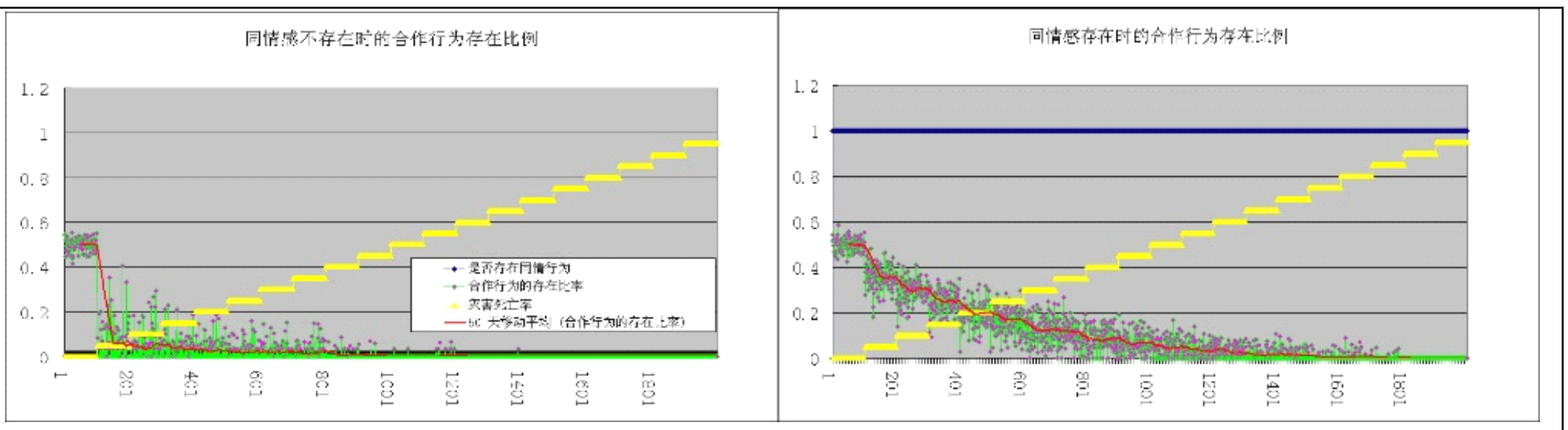

(a) (b)

图一: 上图是在不同的自然选择淘汰率φ下, 对比有无同情行为时, 群体中合作偏好存在的比例。坐标的纵向是演化均衡时合作行为存在的百分比, 横坐标是对不同选择压力进行扫描的次数。 对于每个φ 值各进行 100 次从头开始的演化, 在达到演化均衡时候进行记录。每次均衡后, C[4]的比例记录并在图上标注。(a) 无同情行为的情况 (b) 有同情行为的情况。其中, 格点边长大小 $N = 20$, 合作收益倍率 $b = 3$. φ步长为5%.每步φ是随机初始化后,按照上述博弈步骤, 达到系统均衡的结果记录 C 的百分比。这里所谓的演化均衡值指 C 的比例在演化过程中保持不变15 次。图中，黄色的是选择压力φ值。绿色点是合作生成的比例的标注。 红色线是绿色点 50 次的平均移动拟合线。 对比二图结果，可以看出，比起(a), (b)图 C 存在可能性大很多。

(二) 同情行为导致合作的涌现

系统初始化为全部 DN(即全部为不合作而且没有同情心)，同样对比有无同情行为下的演化结果。采样图见图二。我们注意到，系统可能在一定时间内被锁定在全部被CM, DM, DN 占据，但是未见被红色的 CN 占据的情况，这说明， <u>没有同情行为的合作系统不是稳定的</u>

系统。为此，我们将系统初始化为 100%的 CN。结果显示，CN 的状态往往很快被 DM 入侵，进而 DN 失去稳定性。相反，100%的 CM 状态却相当稳定，见 S2 图说明；该结果暗示着，高比例的合作且有同情心的社会网络（结构）是稳定的。S2 图(c)显示初始化为 100%DN 状态的稳定性。S2 图(D)显示初始化为 100%DM 状态的变化, 迅速显示为多样性的结果，该结果呈现出一个极为有趣的现象，充满同情心的骗子 DM 的社会网络往往容易演化出高比例的合作且有同情心的世界。对比附图组 S2 和 S3，我们可以得到同样的结论。其中 S2（c）可以示意在 DN 中演化出利他合作的过程，可以看到时间上，C 的出现在 DM 之后，而在空间上，C 容易在 DM 中存续。

(三) 同情行为微观空间效果

这个模型框架属于 Nowak 综述[2]中的所述的网络结构，其连接度是 4。 在没有同情心的情况下, b/c > 4 是合作形成的必要条件。在缺省同情行为的步骤 2 下, 实验系统的测试结果与理论数值相符, 不能形成合作结构。

这个模型的步骤, 和一般的惩罚促成合作的步骤一样。差别是, 用同情替代了惩罚。 同情行为如同游戏规则步骤 2。在同情步骤存在的情况下，b=3 可以形成合作。这个模型的核心特征是, 通过怜悯, 缓冲了局域适存度的差异, 提高了采取 C 策略者的生存率。同时, 同情行为在相隔的 C 中, 实现适存度的流动和局域共享。这使得小范围内的合作行为和同情行为得以共生。如图三所示。同情的微观的效应，也可以通过附录中的 S1 图显示。

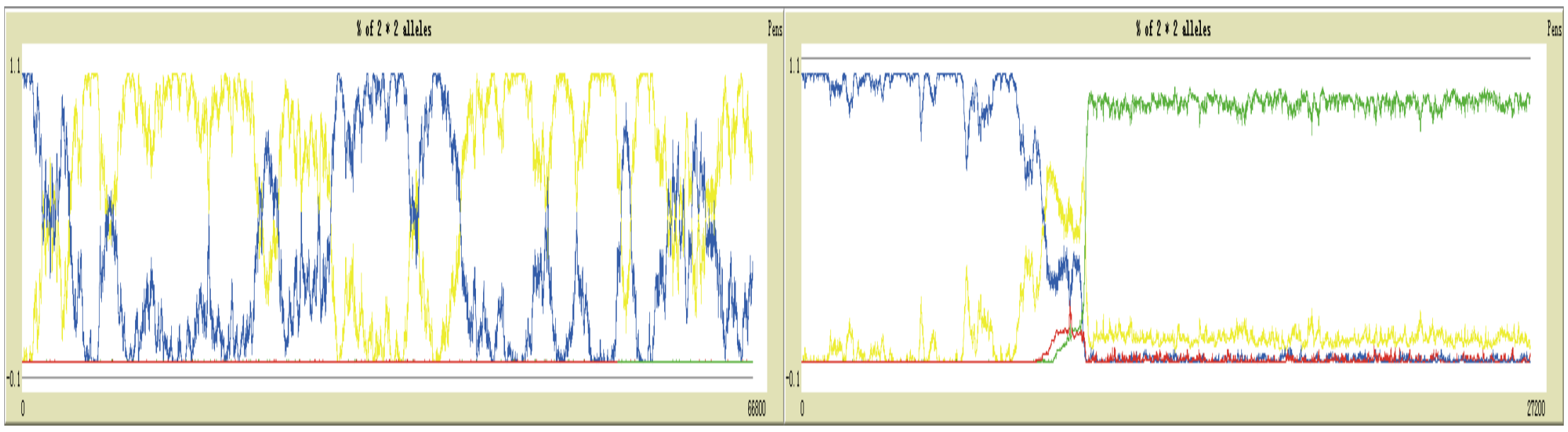

(a) (b)

图二: 有无同情行为的演化过程举例对比。参数为: 变异率为 1/1000, 合作受益倍率为 b = 2.5, 选择压力为 0.2, 同情度为 0.5(完全的爱邻居如同爱自己)。 (a) 同情心不起作用的情况下的结果 (b) 同情心起作用下的结果，亦见图 S2（c)。 坐标的纵向是百分比。 颜色分别是 CM [■], CN [■], DM [■]和 DN [■]。 曲线记录演化过程中的各种行为的比例。

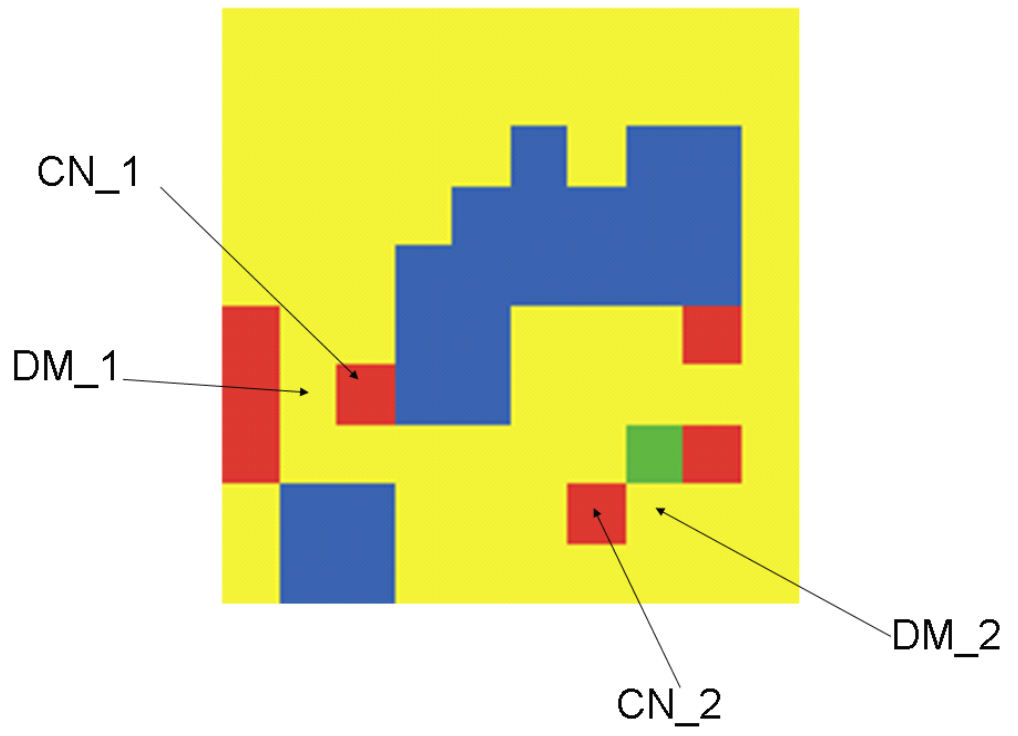

图三: 举例说明同情行为作用的微观效果。 例如, 上图是演化过程的一个截图。 DM_1 是否存在同情行为, 影响了 CN_1 的生存概率。 同样, DM_2 格点的行为影响了 CN_2 的生存概率。 颜色分别代表该格点上的 Agent 的偏好属性, CM [■], CN [■], DM [■]和 DN [■]。

## 结论：

演化博弈的仿真结果明确表明：有同情行为(爱邻居的行为)的系统中，可以涌现稳定的合作结构。合作行为，无论在时间和空间上，都依赖同情行为的存在。同情的存在，同时使得社会网络结构随着合作增益，同情度的不同，显示出 CM [■], CN [■], DM [■]和 DN [■]的共存和多样性。从微观的角度，同情的存在，使得 C 和 DM 可以共存，同时，二个并不近邻的 C 可以共享合作剩余。通过同情微观效果的空间分析，我们注意到，同情的行为如果没有距离的限制，那么其对合作产生的促进效果并不明显（见 S1 图的说明）。一定程度上，本文的结果是对 “爱你的邻居如同爱自己”这一传统箴言的解读。

进一步研究线路可能是: 1)仿真研究怜悯导致合作形成的动力学过程，合作与怜悯度及怜悯范围的依赖关系。 2) 生物学实验研究不同群体的怜悯度和合作偏好的依赖关系; 3) 仿真研究结构同情导致的合作结构稳定性。 4) 同情对不同结构的社会网络的形成及稳定性的影响。 5)在不同的结构和过程中，研究同情的社会影响。同时，对不同选择压力的响应，C 的生成对 M 的动态依赖需要进一步的分析研究。

## 参考文献和说明：

[1] 罗卫东为通讯作者。贡献说明：同情共感导致合作的思想是基于罗卫东的学术判断. 模型框架选择，角色偏好定义与角色行为定义是基于王志坚的学术判断. 仿真模型搭建，仿真实验开展和数据正确性的责任由王志坚承担. 论文由二作者共同完成。本文受“浙江大学 2007 年学科交叉预研基金”的资助，资助的项目名称为“虚拟社会空间的行为规则与公共政策仿真”。

[2] [Nowak2006]Martin A. Nowak, Five Rules for the Evolution of Cooperation, Science, 314: 1560-1563 [DOI: 10.1126/science.1133755]

[3] [Bowles2007] Jung-Kyoo Choi and Samuel Bowles, The Coevolution of Parochial Altruism and War, 318: 636-640 [DOI: 10.1126/science.1144237]; Holly Arrow, The Sharp End of Altruism, Science 26 318: 581-582 [DOI: 10.1126/science.1150316]; Samuel Bowles, Group Competition, Reproductive Leveling, and the Evolution of Human Altruism, Science, 2006 314: 1569-1572 [DOI: 10.1126/science.1134829]

[4] C 的百分比指: 带 C 特征个体数量(CM 和 CN 之和)占整体的数量的比例.

[5][Boyd2007]Robert Boyd and Sarah Mathew, A Narrow Road to Cooperation, Science 29 June 2007 316: 1858-1859 [DOI: 10.1126/science.1144339]; Christoph Hauert, Arne Traulsen, Hannelore Brandt, Martin A. Nowak, and Karl Sigmund, Via Freedom to Coercion: The Emergence of Costly Punishment Science 316: 1905-1907 [DOI: 10.1126/science.1141588]

## 致谢：

作者王志坚感谢高晴的长期支持。二位作者同时感谢汪丁丁教授，叶航教授有益的讨论，感谢 ICSS 博士生贾拥民在成稿过程中的帮助。

## 附件：

附图 S1：同情行为对适存度空间分布的影响

附图 S2：同情行为存在的情况下，100%初始化为(a) CN (b)CM(c)DN(d)DM 的演化例子

附图 S3：同情行为不存在的情况下，100%初始化为(a) CN (b)CM(c)DN(d)DM 的演化例子

本仿真结果在 Netlogo 上实现. (MercyDegree_1). 数据处理采用 MS Excel.

# 附图 S1

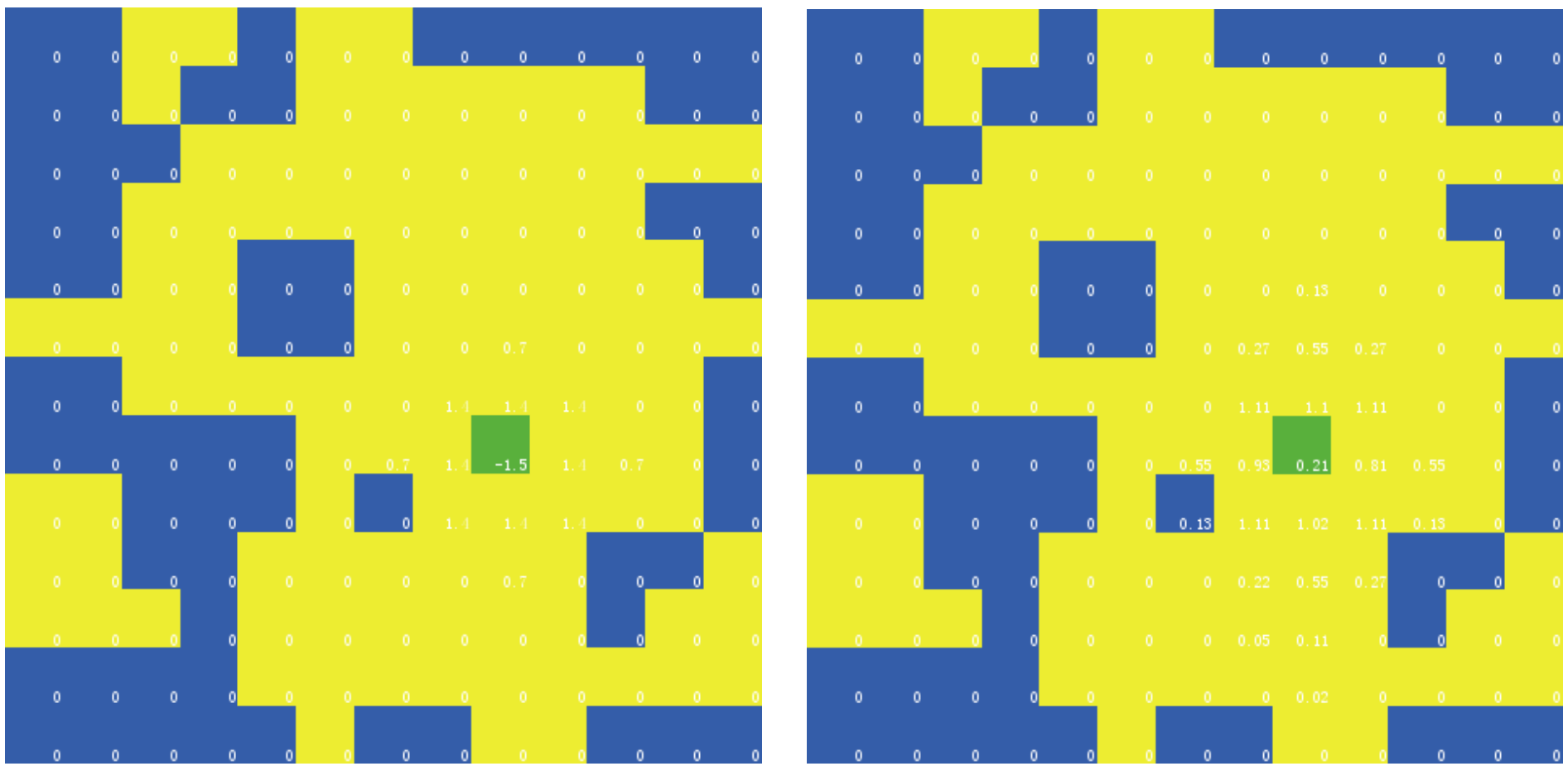

（a） （b）

图 S1：同情行为对适存度空间分布的影响。在同情行为不发生(a)/发生(b)情况下，在 DM 群体中通过变异出现了一个CM 时候，运行游戏规则步骤1，2 的结果。各个格点的Fitness 的数据见图中的数字。由于整体的背景世界是D，那么，在(a)情况下，这个 CM 会被淘汰；而在(b)情况下， 这个 CM 不会被淘汰。 这个选图参数是：每次公共品 C 的支出是 1，合作倍率 3.5，同情度 0.2。同情行为在相隔的C 中，实现适存度的流动和局域共享,在图(b)中不难理解。演化过程中，只要在 fitness 不为 0 的区域中或边界外围，再出现一个带 C 基因的贡献，二个 C 事实上就能共享利他合作行为的效益。同时，同情的行为如果没有距离的限制，那么其对合作产生的促进的效果并不明显，这是因为，围绕着 C 的 fitness 高分布值是 C 的贡献，短程的同情行为是对 C 贡献的反馈性保护。如果同情范围扩大，意味着 C 的贡献得到反馈的机会降低，这样，同情对 C 的保护能力也降低，也就意味着系统 C 和 M 之间的共生加强的能力降低。颜色 CM [■], CN [■], DM [■]和 DN [■]

# 附图 S2

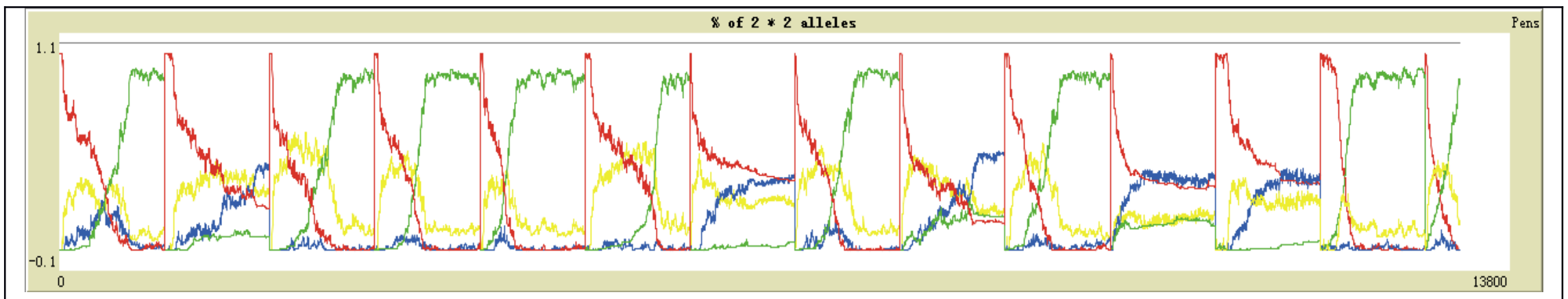

（a）

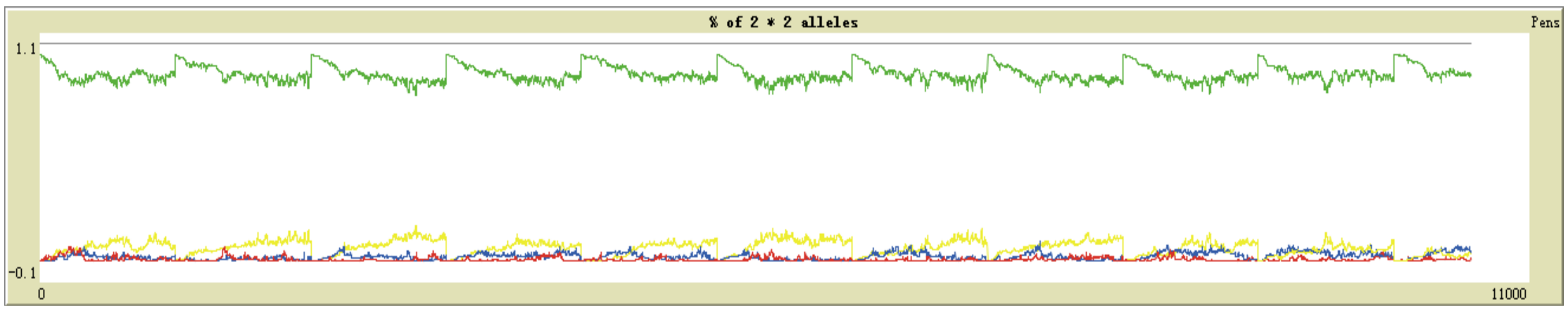

(b)

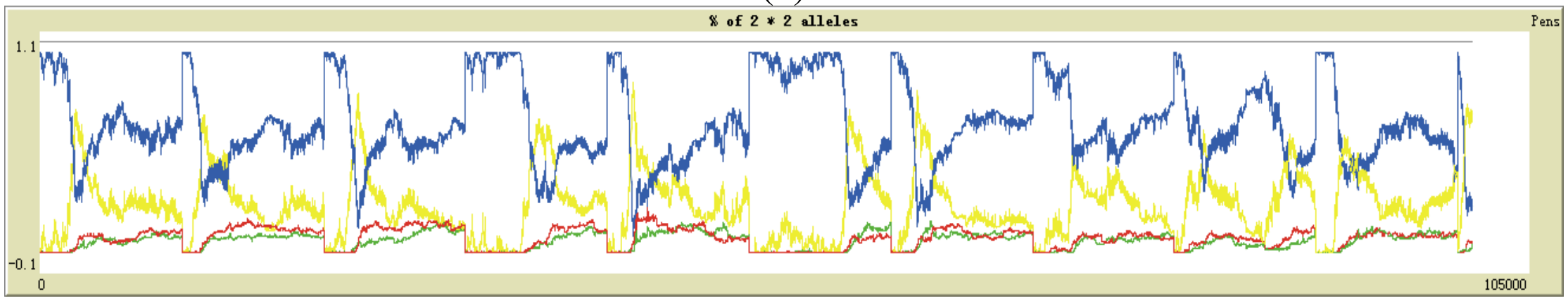

（c）

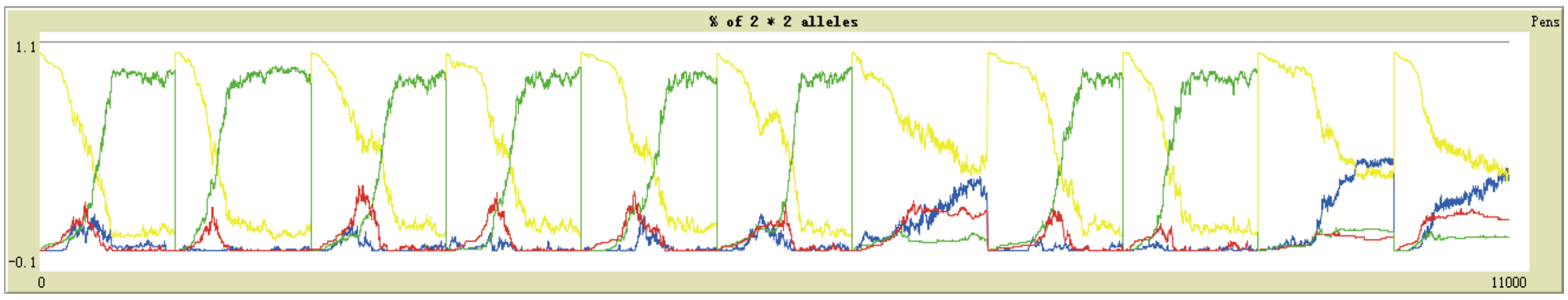

（d）

附图 S2：同情行为存在的情况下，初始化为(a)100%CN(b)100%CM(c)100%DN(d)100%DM 情况下的演化例子。图示为 1000×10 次的结果，其中（c）为 10000×10 次。参数为选择压力 20%，b/c =2.5, 变异率为 0.001，空间大小为 16×16。图(a)显示，初始化为 100%CN 被 DM 侵入的特征。图(b)显示初始化为 100%CM 状态的稳定性。图(c)显示初始化为 100%DN 状态的稳定性。图(D)显示初始化为 100%DM 状态的变化,系统显示为多样性的结果，但是，该结果暗示着，充满同情心的骗子世界往往容易演化出合作且有同情心的世界。坐标的纵向是百分比。 颜色 CM [■], CN [■], DM [■]和 DN [■]。 曲线记录演化过程中的各种行为的比例。

## 附图 S3

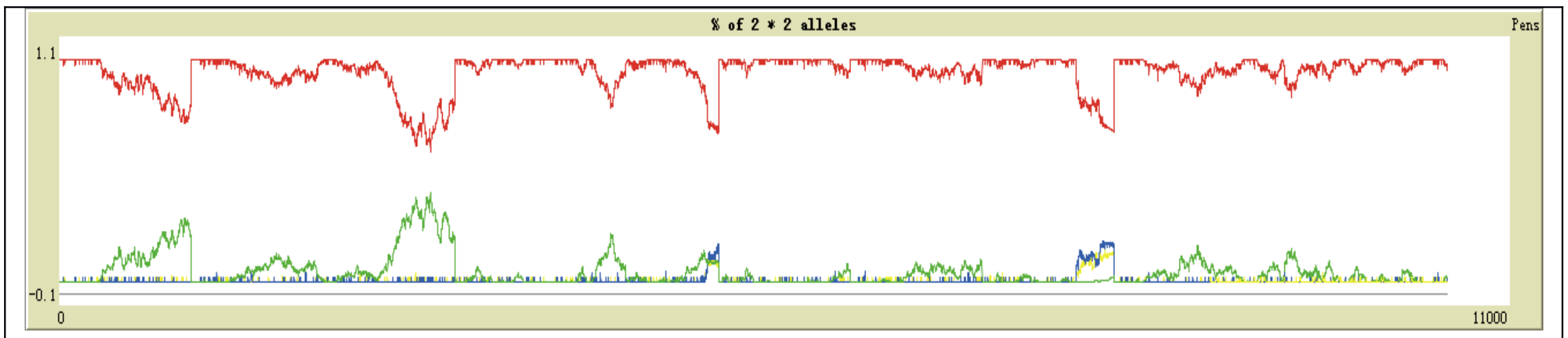

(a)

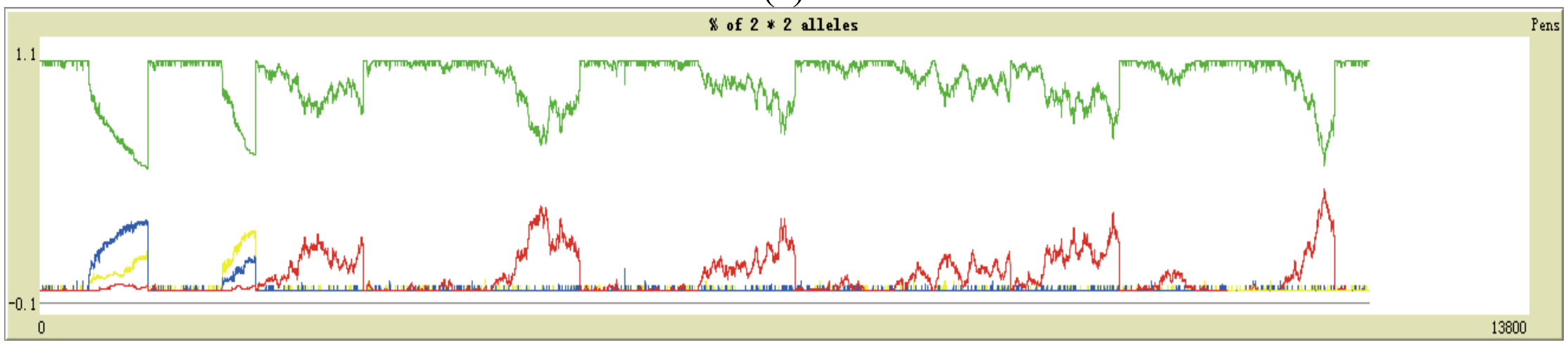

(b)

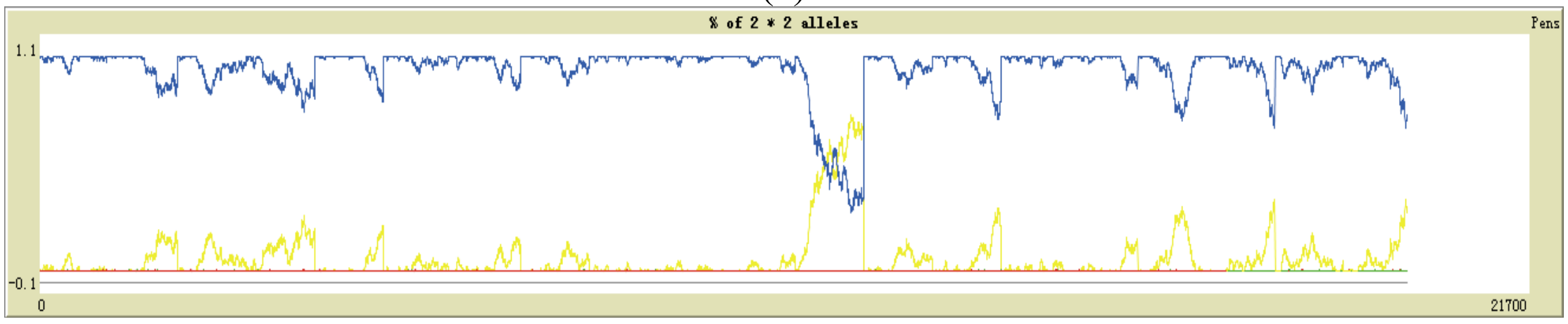

(c)

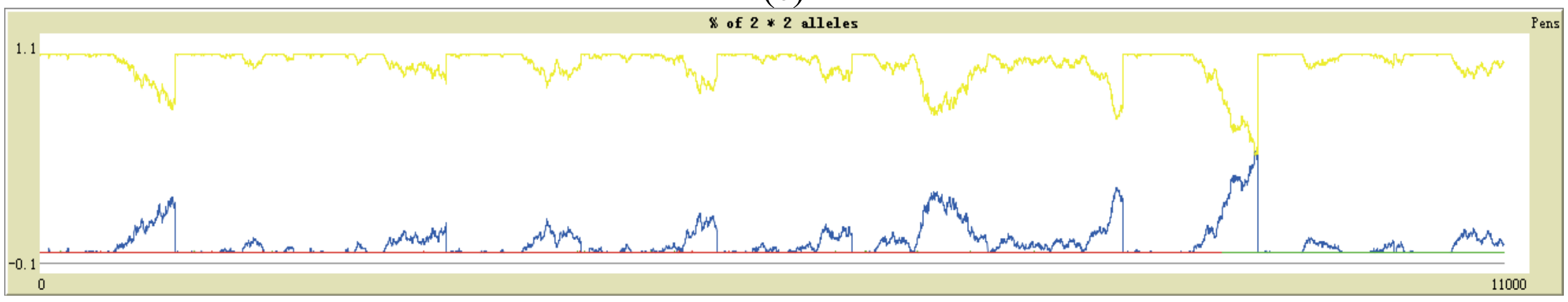

（d）

附图 S3：同情行为不存在的情况下，初始化为(a)100%CN(b)100%CM (c)100%DN (d)100%DM 情况下的演化例子。同情行为外，系统参数与 S2 完全相同。注意，在同情行为不存在时，CM 行为和 CN 行为一样，DM 行为和 DN 行为一样。我们看到，事实上，系统存在二种稳态，一是 100%C， 另外是 100%D。这里 C 的解与本系统空间结构模型和选择模型有关，也与 b/c 有关，也与变异率有关。在 1%的变异率下，100%CN 系统会在 1000 次内退化为 100%D 的世界，而在同情心发挥作用时，高合作比例合作 C 系统在高变异率（如 1%）的变异率下不会退化崩溃为 D 系统。坐标的纵向是百分比。颜色分别是 CM [■], CN [■], DM [■]和 DN [■]。 曲线记录演化过程中的各种行为的比例。